\documentstyle[11pt, aasms4]{article}
\input epsf
\begin{document}
\title{Detection of small scale fluctuations in the near-IR cosmic infrared 
background from long exposure 2MASS fields}

\author{A. Kashlinsky$^1$, S. Odenwald$^2$, J. Mather$^3$, M. F. Skrutskie$^4$ \& 
R. M. Cutri$^5$
\footnote{\scriptsize{
$^1$SSAI, Code 685, Goddard Space Flight Center,
Greenbelt, MD 20771\\
$^2$Raytheon  ITSS, Code 630, Goddard Space
Flight Center, Greenbelt, MD 20771\\
$^3$Code 685, NASA Goddard Space Flight Center, Greenbelt, MD 20771\\
$^4$Dept of Astronomy, University of Virginia, Charlottesville, VA 22903\\
$^5$Infrared Processing and Analysis Center, Pasadena, California
}}
}


\def\plotone#1{\centering \leavevmode
\epsfxsize=\columnwidth \epsfbox{#1}}

\def\wisk#1{\ifmmode{#1}\else{$#1$}\fi}

\def\wm2sr {Wm$^{-2}$sr$^{-1}$ }		
\def\nw2m4sr2 {nW$^2$m$^{-4}$sr$^{-2}$\ }		
\def\nwm2sr {nWm$^{-2}$sr$^{-1}$\ }		
\def\nw2m4sr {nW$^2$m$^{-4}$sr$^{-1}$\ }
\def\Ncut {$N_{\rm cut}$\ }
\def\lt     {\wisk{<}}
\def\gt     {\wisk{>}}
\def\le     {\wisk{_<\atop^=}}
\def\ge     {\wisk{_>\atop^=}}
\def\lsim   {\wisk{_<\atop^{\sim}}}
\def\gsim   {\wisk{_>\atop^{\sim}}}
\def\kms    {\wisk{{\rm ~km~s^{-1}}}}
\def\Lsun   {\wisk{{\rm L_\odot}}}
\def\Msun   {\wisk{{\rm M_\odot}}}
\def\um     { $\mu$m\ }
\def\sig    {\wisk{\sigma}}
\def\etal   {{\sl et~al.\ }}
\def\eg	    {{\it e.g.\ }}
\def\ie     {{\it i.e.\ }}
\def\bsl    {\wisk{\backslash}}
\def\by     {\wisk{\times}}
\def\cosec {\wisk{\rm cosec}}
\def\mic {\wisk{ \mu{\rm m }}}

\def\amin   {\wisk{^\prime\ }}
\def\asec   {\wisk{^{\prime\prime}\ }}
\def\cc     {\wisk{{\rm cm^{-3}\ }}}
\def\deg     {\wisk{^\circ}}
\def\ddeg   {\wisk{{\rlap.}^\circ}}
\def\damin  {\wisk{{\rlap.}^\prime}}
\def\dasec  {\wisk{{\rlap.}^{\prime\prime}}}
\def\approxeq{$\sim \over =$}
\def\abouteq{$\sim \over -$}
\def\percm{cm$^{-1}$}
\def\percmsq{cm$^{-2}$}
\def\percmcub{cm$^{-3}$}
\def\perhz{Hz$^{-1}$}
\def\perpc{$\rm pc^{-1}$}
\def\persec{s$^{-1}$}
\def\peryr{yr$^{-1}$}
\def\te{$\rm T_e$}
\def\tenup#1{10$^{#1}$}
\def\to{\wisk{\rightarrow}}
\def\thin{\thinspace}
\def\uk{$\rm \mu K$}
\def\p{\vskip 13pt}


\begin{abstract}

We report first results for the cosmic infrared background (CIB) 
fluctuations at 1.25, 1.65 and 2.17 $\mu$m obtained from long exposures constructed from 2MASS standard star fields. We have co-added and 
analyzed scans from one such field with a total exposure time $>$ 1 hour, and removed 
sources and other artifacts. The stars and galaxies were 
clipped out to K$_s\simeq$19$^m$ leaving only high-$z$ galaxies (or 
possibly local low-surface-brightness systems). The residual component 
of the diffuse emission on scales from a few arc-sec to a 
few arc-min has a power-law slope consistent with emission produced by clustered galaxies. The noise (and residual artifacts) contribution to the signal is small and the colors of the signal are very different from Galactic stars or air-glow. We therefore identify the signal as CIB fluctuations from the faint unresolved 
galaxies. We show that the present-day galaxies with no evolution would 
produce a significant deficit in the observed CIB fluctuations. Thus the 
dominant contribution to the observed signal must come from high $z$ and 
may indicate high rates of star formation at those epochs.
\end{abstract}

\newpage

\centerline{\bf 1. Introduction}
The near-IR cosmic infrared background (CIB) ($\lambda \lsim 3\mu$m) 
arises mainly from the stellar component of galaxies and probes galactic star formation and evolution at early times. 
The recent surprising, and 
mutually consistent, discoveries of the near-IR CIB from 
the COBE/DIRBE and Japan's IRTS datasets (see Hauser \& Dwek 2001 for review) 
indicated the large amplitude of both 
the CIB fluctuations (Kashlinsky \& Odenwald 2000a - hereafter KO, Matsumoto et 
al 2000,2002) and 
DC levels (Dwek \& Arendt 1998, Gorjian \& Wright 2000, Wright \& Reese 2000, 
Cambresy et al 2001). 
It is hard to explain these measurements by the {\it observed} galaxy 
populations; e.g. at 2.2 \um (K band), the 
faintest observed galaxies give only 30-40\% of the observed CIB  
(Kashlinsky \& Odenwald 2000b, Madau \& Pozzetti 2000). If these detections 
are true, and barring the possible existence of local undetected low surface 
brightness galaxies, the deficit CIB must come 
from fainter galaxies at high $z$ and inaccessible to 
telescopic studies until the launch of NGST. 

In this {\rm Letter} we report the first detection of small angular scale 
fluctuations in the near-IR CIB. We used long integration data constructed from 2MASS 
\footnote{\scriptsize{ This publication makes use of data products from the Two Micron All
Sky Survey, which is a joint project of the University of
Massachusetts and the Infrared Processing and Analysis
Center/California Institute of Technology, funded by the National
Aeronautics and Space Administration and the National Science
Foundation.}}
observations of one standard star field (SSF) with a total exposure 
$\gsim$1 hour. The images were co-added to produce a $8.6^\prime \times 1^\circ$ field from which we removed resolved sources (both stars and distant 
galaxies) and other artifacts as described in Sec.2 and in more 
detail in 
the companion paper (Odenwald, Kashlinsky, Mather, Skrutskie \& Cutri 2002 - 
OKMSC). We then computed 
the fluctuation spectrum of the residual diffuse emission; its slope in the co-added images is consistent with that of galaxy 
clustering. We estimated the contributions to the power spectrum from 
atmospheric fluctuations (AF), remaining Galactic stars (GS) and cirrus 
emission, zodiacal light (ZL), instrument noise (IN) and extinction. These components have different slopes  
and negligible amplitudes compared to the detected signal, suggesting 
that the diffuse component fluctuations in the final image are  dominated by the CIB. If this identification is correct, the 
levels of the CIB from remaining faint (K$_s>$18.5$^m$-19$^m$), and likely high-$z$, galaxies may imply significant star formation at early times. 

\centerline{\bf 2. Data analysis}
2MASS 
conducted measurements with a 3-band camera observing the 
sky simultaneously at J, H and K$_s$ (1.25, 1.65 and 2.17 $\mu$m). The  2MASS Atlas images have 7.8 sec exposure per pixel and are interpolated with 1$^{\prime\prime}$ pixels. 
Our selected SSF, 90565, centered on the star P565-C has 
Galactic and Ecliptic coordinates (243$^\circ , 27^\circ$) and ($21^\circ, 35^\circ$) respectively. The data analyzed here consisted of 2080 
calibrated 8.6$^\prime\times$15$^\prime$ images covering an  $8.6^\prime \times 1^\circ$ swath oriented north-south, obtained during observations at CTIO between Apr and Aug 1998. Calibration, co-adding and other details are 
given in the companion OKMSC paper.
The final image had a total 
exposure $\sim$3,700 secs and the field was divided into 
seven square patches $512^{\prime\prime}$ on the side. 

In each patch, 
individual stars and galaxies were removed by an iterative procedure 
where each pixel with surface brightness exceeding 3 standard deviations for the patch was excised along with 8 neighboring pixels. 
We computed the integrated magnitude limits of the removed sources from the surface brightness threshold. The peak surface brightness of unresolved sources was measured to be +2.5$^m$ 
fainter than the corresponding integrated flux reflecting the part of the beam 
response function within the central 1$^{\prime\prime}$ 2MASS pixel. Because of the variation in background level and associated
noise, the patches were clipped to different point source absolute flux
levels. In K$_s$ band these vary from 
18.5$^m$ to 19$^m$. 

The clipping left $>$90\% of the pixels in the patches, which 
provided a good basis for a Fourier analysis of the diffuse light. Each patch 
was Fourier transformed with pixels weighted by the number of observations. The final results are the same without the weighting. 
The blanked out pixels were assigned flux fluctuation $\delta F(\vec{\theta})$=0; i.e. no power 
is added to the diffuse light emission. After the initial FT of the clipped data, we removed the occasional stripe artifacts from 
diffraction and array row saturation by bright stars. These features produce excess power concentrated along each 
Fourier space axis. The stripes were handled by 
removing a narrow strip of pixels (2 pixels sufficed in practice) in 
the Fourier plane along each axis. The final 3-band color images can be viewed at http://www.gsfc.nasa.gov/gsfc/spacesci/pictures/2mass/2mass.jpg.

The clipped de-striped data have a very different 
distribution than AF. Fig. 1 shows  
$\sqrt{q^2P(q)/2\pi}$, a 
typical fluctuation on angular scale $\theta\!\sim \!\pi/q$. The power spectra follow a power law $P(q) \propto q^n$ with $n\!\sim\! 
-1$, as expected from sources clustered like galaxies. 
For a single standard exposure (7.8 sec) the diffuse emission of the clipped patches has white noise power spectrum from remaining GS and AF multiplied by the Gaussian beam of $2.5^{\prime\prime}$ FWHM. The variance of the clipped maps decreases $\propto$1/$t$ as expected for AF and the plotted power spectrum of AF in the final images was computed from the 7.8 sec image and rescaled to the total exposure time. AF have a white 
noise distribution ($ \sqrt{q^2P(q)/2\pi} \!\propto \!q$) and on the beam scale are 
typically $\sim$20 \nwm2sr . The long 
exposure {\it non-clipped} patches also have a white noise power spectrum from GS and AF. 

\centerline{3. \bf Fluctuations from CIB compared to other terms}
The non-cosmological contributions to the final $P(q)$ are small and 
are as follows:

$\bullet$ {\bf Instrument noise}: This was estimated from  difference maps
for time-ordered string of measurements, each 7.8 sec long, in which
alternating even and odd images are subtracted.
The difference maps show a near white-noise power spectrum and the 
fluctuation in the difference map patches is $< $10\% of the clipped co-added
map. This means that the IN
and time-dependent AF are small and in addition have 
a different power spectrum than the detected signal. To check for stray light effects, we selected from the standard 2MASS product at 7.8 sec exposure per pixel several regions in the Galactic plane with comparable number of stars to our field and processed these with the same clipping algorithm. The clipped power spectra are almost perfectly flat suggesting that stray light does not explain the slopes of our measured 
fluctuation power spectra.

$\bullet$ {\bf Zodiacal light}: Our field is above the 
asteroidal bands so only very small fluctuations in the ZL  
are expected (Kelsall et al 1998). 
Studies of the ZL arcminute-scale structure (Abraham et al 1998) with the ISO/ISOPHOT 
show that at 25$\mu$m no structure can be detected above 0.2\% of the
total ZL brightness. 
The 2MASS observations span a period from
April to August and a solar elongation range from  $115^\circ$ to
$128^\circ$, so any persistent ZL structure would become
uncorrelated during the data averaging process.
The DIRBE based ZL model (Kelsall et al 1998) at Solar elongation of 
120\deg gives the mean ZL intensity of 136 \nwm2sr in 
K$_s$ band and 400 \nwm2sr in J band. 
Based on the limit set by the structure at 25 \um we expect that ZL structure 
will contribute flux fluctuation $<$0.3 \nwm2sr on 
arc-minute scales in the 2MASS bands. 

$\bullet$ {\bf Remaining Galactic stars}: GS brighter than $m_{\rm cut}(K_s) 
\simeq $18.5-19$^m$ were clipped out. Non-clipped patches show that 
GS have a white noise spatial distribution, so over the 2MASS beam $\omega_{\rm beam}$ the flux 
fluctuation, $\sigma_*$, from remaining GS is 
$\sigma_*^2$=$\omega_{\rm beam}^{-1} \int_{m_{\rm cut}}^\infty F_\nu^2(m) n(m) 
dm$, where $n(m) dm$ is the number 
of GS within $dm$ and $F_\nu(m)$=$F_{\nu,0} 
10^{-0.4m}$ is the flux density of a star of magnitude $m$. Using the DIRBE based Galaxy Faint Source Model (Arendt et al 1998) normalized to 2MASS we find that GS fainter 
than $m_{\rm cut}\!\geq $18.5 contribute $\sigma_*\!<$9 \nwm2sr . This component adds in quadrature to the total 
dispersion, contributing $\lsim $5 \% of the total signal in Fig. 
1. 
Stars are not clipped out completely and we checked by increasing the clipping mask size and using other clipping methods, that residual stellar artifacts do not mimic the observed signal. 

$\bullet$ {\bf Galactic cirrus}: From 
COBE/DIRBE data Arendt et al (1998, Table 4) estimate that the 
cirrus flux at 3.5 \um is $\sim$5\% of that at 100 $\mu$m. 
KO  find the fluctuation in the DIRBE maps with 0.3$^{\rm 
o}$ pixels in the (mainly) cirrus emission at 100 \um in the direction of P565-C to be $<$10 \nwm2sr (see their Fig.22). Wright (1998) and KO show that the 
DIRBE data on the Galactic cirrus spatial distribution give a spectral index 
$n\sim -2$, consistent with the IRAS analysis of Gautier et al (1992), but very 
different from the slope of the signal we detect in 
the final co-adds. Combining these results and assuming the 3.5 \um flux to 
represent an upper limit on the  emission at shorter 2MASS  wavelengths 
gives $\sigma_{\rm cirrus}$ well below 1 \nwm2sr .

$\bullet$ {\bf Galactic extinction}:  
The amount of extinction varies strongly
with wavelength: the strongest fluctuations caused by a clumpy interstellar 
medium would
occur in J-band, and with declining strength in H and K$_s$ bands. No such
wavelength-dependent variation in the power spectra in Fig. 1 is apparent. 
Furthermore, at K$_s$ band, the DIRBE based 
models for the amount of interstellar extinction in the near-IR indicate $A 
\simeq 0.05^m$ (Arendt et al 1998) and a
 differential extinction across our $1^\circ$ field of $\approx 0.005^m$. This 
corresponds to an intensity gradient of 0.5\% at the $1^\circ$-scale or less 
than about 0.3 \nwm2sr contribution to the total dispersion.

$\bullet$ {\bf Atmospheric fluctuations}:
Stationary AF (or gradients) should correlate with the air 
column density along the line-of-sight. If important, these effects would 
introduce airmass-dependent correlations into the data during the five months of observations. We computed the flux dispersion in the  individual
scans finding no statistically significant
correlation between the airmass and the flux dispersion. 
AF dominate the diffuse emission for 7.8 sec images and 
integrate down $\propto t^{-1}$, contributing a small fraction to the overall fluctuations (Fig. 1). 

$\bullet$ {\bf Color}: in Fig. 2a the colors of the diffuse signal in the 
final images are compared with those of the 2MASS stars, the 
present day galaxies and the atmosphere. The various 
foreground components and the nearby galaxies have different colors from the signal in Fig. 1.

This discussion suggests that the CIB is the most likely source of the signal in Fig. 1. The largest angular scales probed here correspond to $<$1$h^{-1}$Mpc where galaxy clustering is non-linear and the slope of the power spectrum is consistent with that from galaxy clustering. 

\centerline{4. \bf Discussion}
Assuming that the fluctuations in Fig. 1 are from the CIB, can 
they be produced by the observed non-evolving galaxy populations, or do they require high $z$ evolution? The CIB flux from evolving galaxies is given by $\frac{dF}{dz}$=$\frac{c}{4\pi} \frac{dt}{dz} \frac{{\cal L}_\nu(\nu 
(1+z), z)}{1+z}$, where ${\cal L}_\nu$ is the co-moving 
luminosity density from galaxies. As in Yoshii \& Takahara (1988) we separate the $z$-dependence in ${\cal L}_\nu(z)$ into terms due to K-correction (${\cal 
K_\nu}$), pure luminosity 
evolution (${\cal E_\nu}$) and pure number density evolution (${\cal N_\nu}$), 
i.e. ${\cal L}_{\nu}(z)$=$10^{-0.4[{\cal K_\nu}(z) + {\cal 
E_\nu}(z)+{\cal N_\nu}(z)]} {\cal L}_{\nu}(0)$. We use the K-band for 
quantitative estimates and consider for now only the 
K-correction effects from the observed galaxies. The K-correction was taken from Fig. 1 of JK99. The luminosity 
density was normalized at $z$=0 
using the present-day K-band luminosity function (Gardner et al 1997; Loveday 2000). At small angles, the CIB power spectrum is related to 
$dF/dz$ and the evolving 3-D power spectrum of galaxy clustering, $P_3(k)$ via (e.g. KO) $P(q)$=$\int_0^\infty
\left(\frac{dF}{dz}\right)^2 \frac{1}{c\frac{dt}{dz} d_A^2(z)}
P_3(qd_A^{-1}(z); z) dz$,
where $d_A(z)$ is the angular diameter distance. 

Fig. 2b shows that the total CIB from the observed galaxies 
(K$\lsim$24) is $\simeq$8-10 \nwm2sr (Gardner 1996, Madau \& Pozzetti 2000, 
Kashlinsky \& Odenwald 2000b), a factor of 2-3 smaller than the CIB levels of $F_{\rm K, total}\sim$25 \nwm2sr 
(Matsumoto et al 2000,2002, Gorjian \& Wright 2000, Wright \& Reese 2000). The 
deficit is similarly large at 1.25 \um (Cambresy et al 2001) and 3.5 \um (Dwek 
\& Arendt 1998) and in the $\sim$degree scale 
fluctuation analysis of DIRBE data (KO). The vertical lines show the limiting K 
magnitude, ${\rm K_{cut}}$ of excised galaxies. At $z$=1, the 
angular scale of $1\dasec$  
subtends 10-15$h^{-1}$Kpc, depending on cosmological parameters, so distant galaxies should have been excised almost completely by our 
clipping method. Galaxies up to the clipping threshold contribute only $\sim$5 \nwm2sr to the total CIB. Hence, if the detections of the CIB 
are true the bulk of the flux must be emitted by much fainter galaxies.

Fig. 2c shows the rate of CIB production, $dF/dz$, assuming only 
non-evolving galaxies that were clipped out to ${\rm K_{cut}}$. The K-correction is 
from Fig. 1 of JK99 and no-evolution is a fair assumption for the low-$z$ part of the figure (De Propis et al 1999). Because the brighter galaxies are removed, the 
bulk of the remaining CIB comes from high $z$ probing galaxies at early times. For 
comparison the median observed redshift of K$\sim 19^m$ galaxies is $z\;\sim$0.7-0.8 (Cowie et al 1996). 
We assume 
that all faint galaxies are described by a Schechter-type luminosity 
function and hence are located at cosmological redshifts, but a 
possibility exists that 
at least some of the very faint galaxies are local low-surface brightness 
systems that can not be detected with the current techniques. 
From the K-correction evolution only for $(\Omega, \Omega_\Lambda)$=(0.3,0.7),(1, 0) and (0.3,0) we get that the total flux for 24$>$K$> $19 is 3.7, 2.1 and 2.8 \nwm2sr which compares well with Fig. 2b. 
Far-IR CIB studies from ISO also suggest strong evolution of high-$z$ galaxy populations (Elbaz et al. 1999, Dole et al 2001). 
Hence, if the DC measurements of the K-band CIB are correct, the excess flux must come from fainter and evolving galaxies likely at higher $z$. 

Fig. 2d shows the fluctuation at $1^{\prime\prime}$ vs 
the magnitude $m_{\rm cut}$ of the clipped galaxies for the 2MASS bands. 
Because clipping was done before de-striping the value of $\sigma$ 
shown is {\it before} de-striping. The high numbers for $P(q)$ are consistent with other findings (KO, 
Matsumoto et al 2002) after accounting for the beam difference and the fact that in the large-beam DIRBE and IRTS studies no galaxies were removed. Fig. 1 shows the power 
spectrum of the diffuse CIB from IRTS (Matsumoto et al 2002) and the open 
diamond with error bar shows the fluctuation at $\sim 0.5^\circ$ from the 
COBE DIRBE data (KO). The contribution from galaxies clipped out to K$_s$=18.5$^m$  
can be estimated from the JK99 modeling and gives $\delta F_{\rm rms} \sim$(2-3), (1-1.5) \nwm2sr in J, K bands. When added to the 2MASS numbers 
extrapolated to the appropriate angular scale using the slope of Patch 2 this gives the total (J,K) CIB fluctuation at $0.5^\circ$ of $\sim$(10-12, 2.5-3) \nwm2sr consistent with the KO results.
The largest angular 
scales probed by our analysis after de-striping are $\pi/q 
\sim 1.5^\prime$ which at $z$=1 corresponds to $\simeq$1$h^{-1}$Mpc, where galaxy clustering is 
non-linear. Hence in Fig. 2d we evaluated the expected CIB fluctuations 
assuming only K-correction evolution and two extremes of galaxy clustering evolution: clustering stable in 
proper (thick solid line) and co-moving coordinates. 
The numbers are for $(\Omega, \Omega_\Lambda)$=(0.3,0.7), but models with $\Omega_\Lambda$=0 would give smaller numbers. The deficit in the amplitude of the CIB fluctuation is of a similar factor as the DC component and likely implies strong galaxy evolution at high $z$.

The slope of the detected power spectrum depends on the magnitude cutoff as 
shown in Fig. 2e. If progressively fainter galaxies are removed the slope 
of the power spectrum flattens. This is consistent with fainter galaxies being at higher $z$ when the clustering pattern was less evolved. The 
slope of this dependence is similar in all three bands. The power spectrum of the present-day galaxy clustering on non-linear scales has $n\simeq$--1.3
(Groth \& Peebles 1974, Maddox et al 1990), so it would appear that the J 
band patch with the lowest $m_{\rm cut}$ probes smaller $z$. In 
CDM models the non-linear 
power spectrum of galaxy clustering evolves toward 
steeper slope (higher $|n|$) at low $z$. This is consistent with the trend in Fig. 2d and it appears that H and K bands probe the farthest galaxy populations. This is supported by the colors: the solid 
line in Fig. 2a shows the evolution due to K-correction alone (with 
spectra from JK99) and assuming that all CIB fluctuations in all three bands are dominated by galaxies at the same $z$, in which case the colors would become 
bluer contrary to our observations. A simple way to make the color redder, shown with the filled circle in Fig. 2a, is if different bands, clipped to their respective $m_{\rm cut}$, probe different $z$ with K band probing the earliest times. 

We thank Rick Arendt for fruitful discussions and 
comments on the manuscript.

{\bf REFERENCES}\\
Abraham, P. et a. 1998,in `The Universe as seen by ISO', eds. P. Cox \& M.F. Kessler,pp 145-148.\\
Arendt, R. et al 1998,Ap.J., 508, 74\\
Cambresy, L. 2001,Ap.J.,555,563\\
Cowie, L. et al. 1996, AJ,112,839\\
Cutri, R.M. (IPAC/Caltech), Skrutskie, M.F. (UMASS), et al (IPAC/Caltech) 
http://www.ipac.caltech.edu/2mass/releases/second/doc/explsup.html\\
De Propis et al 1999, Ap.J.,118,719\\
Djorgovski, S. et al 1995, Ap.J., 438, L13\\
Dole, H. et al 2001, A \& A, 372,364\\
Dwek, E. \& Arendt, R. 1998,Ap.J.,508,L9\\
Elbaz, D. et al 1999, A \& A, 351, L37\\
Gardner, J. 1996, in ``Unveiling the cosmic 
infrared background", ed. E. Dwek, p. 127\\
Gardner, J.P. 1997, Ap.J.,480,L99\\
Gautier, D. 1992,AJ,103,1313\\
Gorjian, V., Wright, E.L. \& Chary, R.R. 2000,Ap.J.,536,550\\
Groth, E. \& Peebles, P.J.E. 1977,Ap.J.,217,385\\
Hauser, M. \& Dwek, E. 2001,ARAA,39,249\\
Jimenez, R. \& Kashlinsky, A. 1999,Ap.J.,511,16 (JK99)\\
Kashlinsky \& Odenwald 2000a, Ap.J., 528, 74\\
Kashlinsky \& Odenwald 2000b, Science, 289,246\\
Kelsall, T. et al 1998,Ap.J.,508,44\\
Loveday, J. 2000,MNRAS,312,L281\\
Madau, P. \& Pozzetti, L. 2000,MNRAS,312,L9\\
Matsumoto, M. et al 2000,in ``ISO surveys of a dusty Universe", eds. Lemke, D. et al. p.96.\\
Matsumoto, M. et al 2002, Ap.J., submitted.\\
Odenwald, Kashlinsky, Mather, Skrutskie and Cutri 2002, Ap.J.,submitted - 
OKMSC\\
Wright, E.L. 1998,Ap.J.,496,1\\
Wright, E.L. \& Reese, E.D. 2000,Ap.J.,545,43\\
Yoshii, Y. \& Takahara, F. 1988,Ap.J.,326,1\\

\newpage
\centerline{\bf FIGURE CAPTIONS}

Fig. 1 - The power spectrum of the diffuse emission in the final images is 
shown with crosses with errors. Patch number is shown in the upper left corner 
of the top panels. The number in the upper right corner of each panels shows the final $\sigma$ {\it after} de-striping. Open diamonds with 92\% errors show the fluctuation on larger angular scale from KO analysis of COBE DIRBE data. Thick 
solid line shows the CIB result from IRTS analysis (Matsumoto et al 2002). 
Filled small diamonds show the atmospheric contribution evaluated from 7.8 sec 
exposure and divided by the average number of co-adds in the patch. The number 
in the lower right corner gives the limiting magnitude for removing sources from the clipped patch.

Fig. 2 - {\bf (a)} H--K vs J--K colors. Crosses correspond to fluctuations 
in the final 2MASS images shown in Fig. 1. Square is the color of the 
atmosphere. Ellipse corresponds to the locus of $\sim 10^5$ high latitude  Galactic field stars from 2MASS data. 
Plus denotes the position of the present day galaxies using SED from JK99.
Solid line show the evolution of colors out to $z$=2 using K-correction and SED 
from JK99. Filled circle shows an example of colors from high-$z$ galaxies with 
SED from JK99 assuming that they are at different $z$: $z$=0.7 in J, 0.5 in H 
and 2 in K bands. 
{\bf (b)} Cumulative flux distribution as function of K magnitude from galaxies 
from deep galaxy surveys (e.g. Djorgovski et al 1995). Vertical lines denote the range of limiting magnitudes of clipped point sources in the seven patches. 
{\bf (c)} Rate of CIB flux production vs $z$ assuming only K-correction in the 
present-day population of galaxies. SED was taken from JK99. Solid, dashed and 
dotted lines are for $(\Omega, \Omega_\Lambda)$ = (0.3,0.7), (0.3,0) and (1,0) 
respectively. Thin lines of each type correspond to the largest clipping 
magnitude galaxies (K=18.6 in Patch 2) and thick lines to the faintest (K=19 in 
Patch 3).
{\bf (d)} Amplitude of the rms CIB flux fluctuation from Fig. 1 vs $\sigma$ 
{\it before} de-striping. The magnitude of the clipped out galaxies (and stars), $(m_J, m_H, m_K)$=(24.3,23.5,22.8)--$2.5 \log 
\sigma$,  
is shown on the top horizontal axis, where $A_m = 1.6, 0.8, 0$ in J, H, K bands 
respectively. Triangles, squares and crosses correspond to J, H, K bands. 
{\bf (e)} Power law index $n$ for the final power spectrum in Fig. 1 is plotted 
vs the $\sigma$ before de-striping. Same symbol notation as in (d).

\clearpage
\begin{figure}
\centering
\leavevmode
\epsfxsize=1.0
\columnwidth
\epsfbox{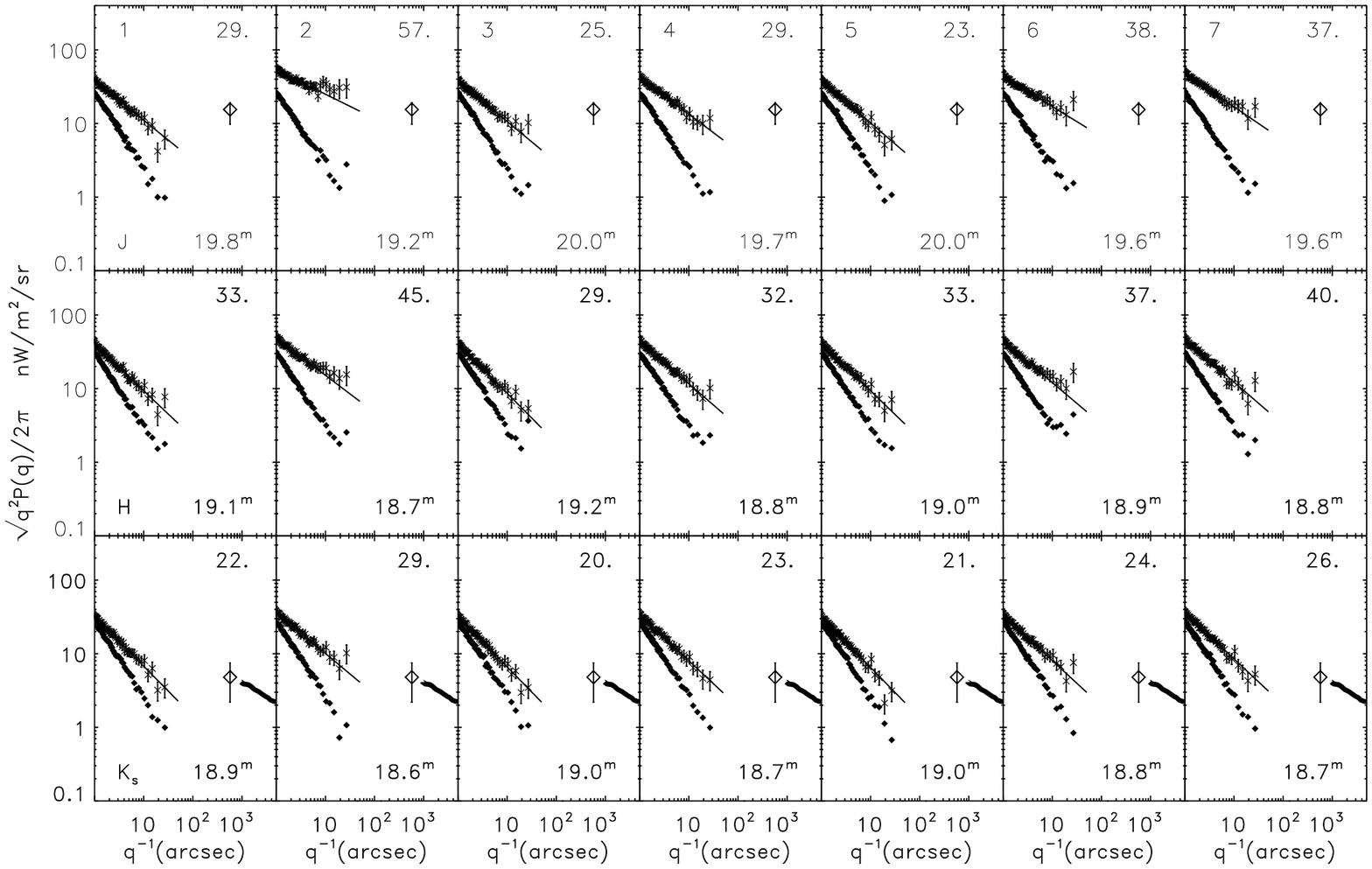}
\caption[]{
{\scriptsize{}
}
}
\end{figure}

\clearpage
\begin{figure}
\centering
\leavevmode
\epsfxsize=1.0
\columnwidth
\epsfbox{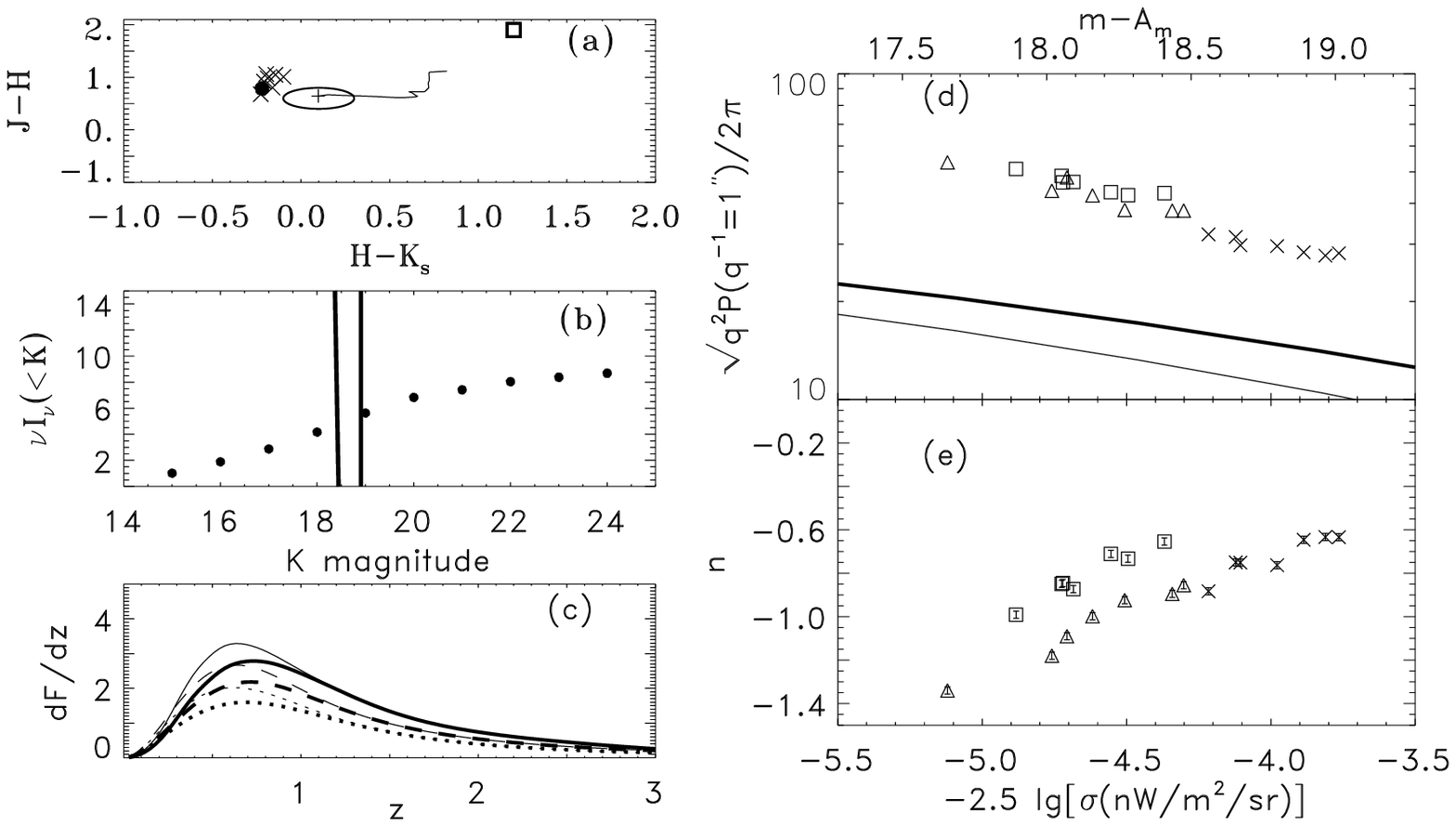}
\caption[]{
{\scriptsize{} 
}
}
\end{figure}

\end{document}